\documentstyle[aps]{revtex}
\begin{document}
\title{New mechanism of first order magnetization process in multisublattice
rare-earth compounds}
\author{V. Yu. Irkhin$^{*}$}
\address{Institute of Metal Physics, Kovalevskay str. 18, 620219 Ekaterinburg, Russia}
\maketitle

\begin{abstract}
First-order field-induced spin reorientation transitions in multisublattice
intermetallic compounds are considered within an anisotropic Heisenberg
model. Unlike previous works, only leading-order anisotropy constants of the
sublattices are included. The exchange interactions both between the
rare-earth (RE) and transition metal (TM) sublattices and within the RE
sublattice play an important role in the model. The latter interaction
results in the occurrence of metastable states. One of the minima
corresponds to the non-collinear magnetic structure where the RE subsystem
is divided into two sublattices, and their moments deviate in opposite
directions from the TM moment direction. The possible FOMP scenarios for R$%
_2 $Fe$_{17}$, HoFe$_{11}$Ti, and TbMn$_6$Sn$_6$ systems are discussed.
\end{abstract}


\section{Introduction}

The phenomenon of FOMP (First Order Magnetization Process) was discovered
during last years in many intermetallic compounds. Here belong a number of
practically important compounds of transition metals with rare-earths (RE)
like R$_2$Fe$_{14}$B \cite{herbst,yamada}, Nd$_2$Co$_{14}$B \cite{ndco}, R$%
_2 $Fe$_{17}$ \cite{dy17,tb17,er17,r17,park}, RFe$_{11}$Ti \cite
{smfe11,hofe11,nik1}, RMn$_6$Sn$_6$ \cite{zajkov}, Sm$_{1-x}$Nd$_x$Co$_5$ 
\cite{erm,ndco5}, and also some ferrites \cite{ferr}. In some directions,
magnetization jumps with increasing external magnetic field are observed in
these systems. Usually this effect is treated phenomenologically as a result
of competition of magnetic anisotropy (MA) constants $K^{(n)}$ of different
order \cite{asti,pareti}. Presence of large high-order constants enables one
to obtain local minima of free energy and thereby first-order transitions.
However, such an approach meets with a number of difficulties. Large values
of $K^{(n>1)}$ required are not confirmed by crystal-field calculations (see
Refs.\cite{herbst,yamada,zhong}). Besides that, high order $K^{(n)}$ are
proportional to high powers of magnetization and should rapidly decrease
with increasing temperature $T$. This effect should be stronger in the case
where $K^{(n)}$ come mainly from RE sublattice: the rare-earth moments can
become rather small at $T$ well below the Curie temperature $T_C.$ The
reason is that the exchange interaction $J_{{\rm RT}},$ which acts on RE
ions from the transition metal (TM) system, may be considerably smaller than
the interaction $J_{{\rm TT}}$ which determines $T_C$ \cite{ros}. Thus, the
local minima explaining FOMP should exist only at low temperatures. However,
FOMP is observed as a rule in a wide temperature region \cite{r17,zajkov}.
The dependences $K^{(n)}(T)$ fitted from magnetization curves turn out to be
rather complicated and sometimes seem to be artificial.

The data discussed lead to the conclusion that effects of many-sublattice
structure are important in systems demonstrating FOMP \cite{pareti}. It is
known that in a number of intermetallic RE systems (Tm$_2$Fe$_{17},$ RFe$%
_{11}$Ti, TbMn$_6$Sn$_6$) temperature-induced orientational transitions are
caused by competing anisotropy of RE and TM sublattices (in hexaferrites BaZn%
$_{2-x}$Co$_x$Fe$_{16}$O$_{27}$ \cite{ferr}, competition of Fe and Co
sublattices plays a similar role). It is natural to suppose that this factor
is essential for the field-induced transitions too. Although the
many-sublattice case was considered earlier \cite{asti1}, these treatments
were also based on the idea of competion of high-order constants.

In the present work we propose a new mechanism of FOMP in a multisublattice
magnet, which is connected with the formation of a non-collinear spin
structure. The corresponding theoretical model is formulated in Sect.2. In
Sect.3 we consider the FOMP picture in various cases and discuss
experimental situations.

\section{The theoretical model}

We start from the Heisenberg model of uniaxial multisublattice crystal with
inclusion of magnetic anisotropy. The MA constants in the problem are
usually treated as phenomenological parameters which are obtained by fitting
magnetization curves. When describing FOMP one introduces as a rule large
high-order constants, $K^{(3)}\sim K^{(2)}\sim K^{(1)}$. To avoid the
difficulties discussed above, we include only leading-order MA constants for
different magnetic sublattices, $K_i=K_i^{(1)}$. Using the mean-field
approximation we write down the expression for the total energy 
\begin{equation}
E=\sum_{i>j}J_{ij}\cos (\theta _i-\theta _j)+\sum_i(K_i\sin ^2\theta _i-{\bf %
HM}_i).  \label{ham}
\end{equation}
Here $i,j$ are sublattice indices, ${\bf M}_i$ are the corresponding
magnetic moments which make the angles $\theta _i$ with the $z$ axis, $%
J_{ij} $ are exchange parameters, ${\bf H}$ is the external magnetic field.
For the sake of convenience, we have included the dimensionless factors of $%
M_i=\langle {\bf M}_i\rangle $ into $J_{ij},$ so that $J_{ij}\propto M_iM_j.$
The crystal and magnetic structures contain TM and RE sublattices which are
coupled by the exchange interaction $J_{{\rm RT}}$. The TM subsystem yields
usually main contribution to magnetic moment, and the RE one to MA. The most
strong interaction $J_{{\rm TT}}$ plays the role of a mean field in the TM
subsystem and needs not to be included explicitly.

As demonstrate crystal-field considerations, formation of non-collinear
magnetic structures is possible in multisublattice structures, e.g., for the
systems R$_2$Fe$_{14}$B \cite{yamada,herbst}. The simplest case of a
two-sublattice magnet was analyzed earlier in Ref.\cite{ros}. It was shown
that, in the case where $\ K_{{\rm T}}$ and $K_{{\rm R}}\ $have opposite
signs (competing anisotropy), the sublattice moments become non-parallel
provided that

\begin{equation}
|J_{{\rm RT}}(1/K_{{\rm R}}+1/K_{{\rm T}})|<2  \label{crit}
\end{equation}
If the absolute values of $K_{{\rm T}}$ and $K_{{\rm R}}$ are close, this
condition can hold even for rather strong exchange $J_{{\rm RT}}.$

The two-sublattice problem with non-collinear magnetic structure enables one
to obtain a local energy minimum with including magnetic field, but for one
field direction only. To avoid this difficulty, we have to introduce,
instead of one RE subsystem, {\it two} equivalent RE sublattices, RI and
RII, with $K_{{\rm RI}}=K_{{\rm RII}}=K_{{\rm R}}$, $J_{{\rm RIT}}=J_{{\rm %
RIIT}}=J_{{\rm RT}}.$ This model preserves the correct axial symmetry and
yields the energy minimum for both field directions. Of course, for real
compounds with complicated crystal structures each `sublattice' corresponds
to several crystal positions. To stabilize the non-collinear structure with
split RE sublattice we include the antiferromagnetic exchange interaction
between the sublattices, $J_{{\rm RR}}>0$. The $J_{{\rm RT}}$ and $J_{{\rm RR%
}}$ values cannot be reliably obtained from existing experimental data, but
for the long-range RKKY exchange interaction $|J_{{\rm RR}}|$ may be assumed
not small. The essential property of such a model is the presence of
metastable states already at $H=0$. Owing to this fact, first-order phase
transitions are possible which occur at crossing the energy levels in the
external magnetic field.

\section{FOMP scenarios for various experimental situations}

The idea of the FOMP mechanism can be illustrated on a simple toy model
where the TM moment is rigidly fixed by strong MA in the $x$ direction (or
in the $xy$ plane for the three-dimensional problem). The RE ions which
possess more weak easy-axis MA tend to incline to the $z$ direction, each
ion having two local minima. The interaction $J_{{\rm RR}}$ results in that
the degeneracy of the minima is lifted, so that the moments of RE
sublattices deviate in opposite directions (Fig.1a). With increasing
magnetic field along the $z$ axis, a first-order transition occurs to the
configuration where RE moments are almost parallel (Fig.1b). A smooth
transition does not occur because of the influence of MA of RE ions.

Now we pass to a more realistic situation of finite (and even small) MA of
TM subsystem. First we consider the situation where $K_{{\rm R}}>0,K_{{\rm T}%
}<0,$ and the easy-axis MA of RE sublattice predominates. This is the case
for the systems Tm$_2$Fe$_{17}$ and TbMn$_6$Sn$_6,$ as follows from the
existence of easy-axis - easy-plane spin-reorientation transitions with
increasing temperature ($|K_{{\rm R}}|$ decreases more rapidly than $|K_{%
{\rm T}}|$). Two possible metastable states are shown in Fig.2, the exchange 
$J_{{\rm RT}}$ being supposed antiferromagnetic, which corresponds to heavy
rare-earth ions, e.g., to Tb and Tm. In the case where $J_{{\rm RT}}<0,$
which occurs for light rare-earths, e.g., for Nd and Pr, the picture differs
by the opposite direction of RE moments.

In the zero magnetic field the first state (Fig.2a) is realized with the
collinear spin configuration along the $z$ axis (easy direction for the RE
sublattice). At increasing the field ${\bf H}\parallel x$, the magnetization
vector starts to rotate with retaining the nearly antiparallel orientation
of TM and RE moments. For simplicity, we neglect in the analytical treatment
the non-collinearity of TM and RE moments for this state, which is weak due
to large value of $|J_{{\rm RT}}|$. Then the energy in the state (a) is
expressed in terms of the angle $\theta $ beween the TM moment and $z$ axis,

\begin{eqnarray}
E_{{\rm a}} &=&-2|J_{{\rm RT}}|+J_{{\rm RR}}-H(M_{{\rm T}}\pm 2M_{{\rm R}%
})\sin \theta  \nonumber \\
&&\ \ \ \ \ \ \ \ \ +(K_{{\rm T}}+2K_{{\rm R}})\sin ^2\theta  \nonumber \\
\ &=&-2|J_{{\rm RT}}|+J_{{\rm RR}}-\frac{(M_{{\rm T}}\pm 2M_{{\rm R}})^2}{%
4(2K_{{\rm R}}+K_{{\rm T}})}H^2  \label{e1}
\end{eqnarray}
the sign $+(-)$ corresponding to $J_{{\rm RT}}<0$ ($J_{{\rm RT}}>0$). The
second state (Fig.2b) posseses a non-collinear spin configuration with split
RE sublattice owing to the gain in the MA energy for the RE subsystem, and
also due to the interaction $J_{{\rm RR}}$. In this state the spin
configuration depends weakly on magnetic field and the TM moment remains
parallel to $x$ axis, so that 
\begin{eqnarray}
E_{{\rm b}} &=&-2|J_{{\rm RT}}|\cos \alpha +J_{{\rm RR}}\cos 2\alpha 
\nonumber \\
&&\ \ \ \ \ \ \ \ -H(M_{{\rm T}}\pm 2M_{{\rm R}}\cos \alpha )+K_{{\rm T}%
}+2K_{{\rm R}}\cos ^2\alpha  \nonumber \\
\ &=&K_{{\rm T}}-HM_{{\rm T}}-J_{{\rm RR}}-\frac{(J_{{\rm RT}}-HM_{{\rm R}%
})^2}{4(J_{{\rm RR}}+K_{{\rm R}})}  \label{e2}
\end{eqnarray}
where $\alpha $ ($0<\alpha <\pi /2$) is the angle of the RE moment deviation
from the $x$ axis. The total energy in the state (b) has an almost linear
field dependence and decreases more rapidly than that for the state (a) with
rotating moment (Fig.3). The angles $\theta $ and $\alpha $ are obtained by
minimization of Eqs.(\ref{e1}), (\ref{e2}), 
\begin{equation}
\sin \theta =\frac{H(M_{{\rm T}}\pm 2M_{{\rm R}})}{2(2K_{{\rm R}}+K_{{\rm T}%
})},\cos \alpha =\frac{|J_{{\rm RT}}-HM_{{\rm R}}|}{2(J_{{\rm RR}}+K_{{\rm R}%
})},  \label{at1}
\end{equation}
Thus the local minimum (non-collinear state) occurs at 
\begin{equation}
J_{{\rm RR}}+K_{{\rm R}}>|J_{{\rm RT}}-HM_{{\rm R}}|/2.  \label{crr}
\end{equation}
Note that the minimum can occur with including magnetic field, even if it is
a absent at $H=0.$ The transition field and the angle $\alpha $ (but not the
magnetization jump) turn out to be small provided that $2|J_{{\rm RR}}+K_{%
{\rm R}}|\simeq |J_{{\rm RT}}|$. Since $J_{{\rm RR}}$ and $K_{{\rm R}}$ are
proportional to $M^2(T)$ (unlike the constants $K^{(n>1)}$ in standard FOMP
considerations, which decrease rapidly with increasing $T$), the criterion (%
\ref{crr}) depends rather weakly on $T.$ Therefore the mechanism discussed
enables one to obtain FOMP in a wide temperature region.

The numerical calculations were performed by direct minimization of total
energy (1) without approximations discussed above. The moment values were
taken to be $M_{{\rm T}}=4,M_{{\rm RI}}=M_{{\rm RII}}=M_{{\rm R}}/2=1$,
which corresponds roughly to R$_2$Fe$_{17}$ compounds (e.g., for Tb$_2$Fe$%
_{17}$ the moment of Fe subsystem exceeds by two times that of RE subsystem, 
$M_{{\rm Tb}}=18.6\mu _B,M_{{\rm Fe}}=36.6\mu _B$ per formula unit \cite
{tb17}). At some critical field $E_{{\rm a}}$ becomes equal to $E_{{\rm b}}$
(Fig.3), and the magnetization demonstrates a vertical jump (Fig.4). It
should be noted that the transition field is much smaller than the exchange
interaction $J_{{\rm RR}},$ unlike the situation of Fig.1. After the jump,
the magnetization increases rather slowly because of large value of $J_{{\rm %
RT}}.$ On the other hand, the magnetization depends very weakly on the field 
${\bf H}\parallel z$ because of the strong interaction $J_{{\rm RT}}.$ Owing
to the deviation from the antiparallel alignment, the magnetization in the
hard direction $x$ after the transition turns out to be larger than the
zero-field magnetization in the easy direction $z$ ($M_x>M_z=4-2=2$). Such a
behavior is demonstrated by experimental dependences $M(H)$ in a number of
systems, e.g., in Tm$_2$Fe$_{17}$ \cite{r17,park}.

The zero-field canting angle $\alpha $ of the RE sublattice for the
parameters of Fig.3 makes up about 20$^0.$ Such small values do not
contradict likely experimental data. Indications of formation of the
non-collinear structure after the transition are given by neutron scattering
data on Tm$_2$Fe$_{17}$ \cite{park} and TbMn$_6$Sn$_6$ \cite{pirogov}.
Non-collinear structures with the split RE sublattice were also observed for
R$_2$Fe$_{14}$B systems \cite{neut}.

In the system Tb$_2$Fe$_{17}$ \cite{tb17,r17} the Fe sublattice has the
easy-plane magnetic anisotropy, as well as the Tb sublattice (although Tb
ions in different positions are supposed to have MA of different signs).
Thus, the competing anisotropy situation does not occur. Nevertheless, in
this case our approach also enables one to obtain FOMP, but the transition
field turns out to be somewhat larger. Fig.5 shows the result of calculation
of magnetization for the constants of the same sign, $K_{{\rm R}}>0,K_{{\rm T%
}}>0$ (the easy-plane and easy-axis problems are equivalent provided that we
restirict ourselves to the plane model). One can see that before the jump
the field dependence of the magnetization is practically linear, unlike
Fig.4. The state with split RE sublattices occurs only with increasing
magnetic field.

The curves obtained demonstrate a similarity to experimental ones. Let us
try to perform a more detailed comparison with experimental data. The
critical FOMP field is determined by the overall scale of the anisotropy
parameters. Unfortunately, MA constants are usually fitted from
magnetization curves with including high-order constants which contradicts
our model. Therefore we shall concentrate on the cases where the parameters
of the model can be determined from independent experiments. Using the data
for Tm$_2$Fe$_{17}$ from Ref.\cite{park}, $M_{{\rm Tm}}\simeq 12\mu _B,M_{%
{\rm Fe}}\simeq 30\mu _B$ per formula unit, $J_{{\rm TmFe}}=60$J/cm$^3,$ $K_{%
{\rm Fe}}=-2.5$ J/cm$^3,K_{{\rm Tm}}=7$ J/cm$^3,$ we can fit the
experimental FOMP field, $\mu _0H=5$T, by taking $J_{{\rm RR}}/J_{{\rm RT}%
}\simeq 0.5$. Unfortunately, the value of the magnetization jump cannot be
reliably determined from the results of \cite{r17,park}.

The values of MA constants for Tb$_2$Fe$_{17}$ can be estimated from NMR
data \cite{li} (see Ref.\cite{ii}), $K_{{\rm Tb}}=-10$ J/cm$^3,$ $K_{{\rm Tb}%
}=3$ J/cm$^3$ for the positions 2b and 2d, $K_{{\rm Fe}}=-3$ J/cm$^3.$ Of
course, the accuracy of these values is rather poor. After averaging the
values of $K_{{\rm Tb}}$ over sublattices such a relation between MA
constants corresponds roughly to the parameters of Fig.5. The rescaled FOMP
critical field from this Figure turns out to be by about two times larger
than the experimental value which makes up $\mu _0H\simeq 4$T. The agreement
can be made better provided that we take somewhat smaller values of $|K_{%
{\rm Fe}}|$ and increase the ratio $J_{{\rm RR}}/J_{{\rm RT}}.$ Besides
that, the situation in Tb$_2$Fe$_{17}$ is complicated because of existence
of two FOMP-type transitions \cite{r17}.

Now we treat the case where easy-axis anisotropy $K_{{\rm T}}$ dominates
over MA of RE sublattice, which has an opposite sign (such a stituation
takes place, e.g., for the compound HoFe$_{11}$Ti). Then the qualitative
picture of FOMP turns out to be different from Fig.2. As shown in Fig.6a, at 
$H=0$ RE moments deviate from the axis $z$ (easy direction for TM
sublattice) by the angle 
\begin{equation}
\alpha =\arccos \left[ \frac{|J_{{\rm RT}}|}{2(J_{{\rm RR}}-K_{{\rm R}})}%
\right] .  \label{a11}
\end{equation}
This configuration begins to rotate when including the small magnetic field
in the $x$ direction. At further increasing the field, the nearly collinear
spin configuration shown in Fig.6b becomes favorable. As follows from (\ref
{a11}), FOMP is possible at $J_{{\rm RR}}-K_{{\rm R}}>|J_{{\rm RT}}|/2.$
Note that the solution (local minimum), corresponding to the non-collinear
structure, can vanish with increasing $H$ somewhat above the critical field
of FOMP. A structure of the type Fig.6a (but with ferromagnetic interaction $%
J_{{\rm RT}}<0$) was proposed for Y$_{1-x}$Nd$_x$Co$_5$ on the basis of
magnetic measurements \cite{erm1}.

After the jump, the magnetization in the $x$ direction continues to increase
rather rapidly due to rotation of TM moments (Fig.6b), and becomes saturated
only when the moments lie along the $x$ axis ($M_x=2$). Absence of the
saturation immediately after the jump corresponds to FOMP-II according to
the classification \cite{asti,pareti}. The result of magnetization
calculation is shown in Fig.7. For simplicity, we restrict ourselves again
to the plane problem ($xz$ plane model). The parameter values are chosen to
demonstrate existence of the saturation region, although FOMP can be
obtained for smaller $|K_i/J_{{\rm RT}}|$ and at smaller $H$ too. The
dependence obtained is in a qualitative agreement with that for HoFe$_{11}$%
Ti \cite{nik1}. This system is close to instability with respect to the
formation of the `easy-cone' structure \cite{nik1} owing to competing MA of
Ho and Fe sublattices with $K_{{\rm R}}<0,K_{{\rm T}}>0$.

\section{Conclusions}

To conclude, we have proposed a simple model which reproduces qualitatively
the FOMP pictures in some RE systems with different sign combinations of
sublattice MA constants. It should be noted that the field-induced
transitions considered are similar in some respects to the spin-flop
transitions in anisotropic antiferromagnets. In the latter situation,
formation of metastable states and magnetization jumps are possible too
(see, e.g., Ref.\cite{vons}). However, our picture is more complicated since
this includes the TM sublattice.

The consideration performed enables one to avoid some difficulties and
reduce the number of fitting parameters in comparison with the treatments
that use high-order MA constants. Further experimental investigations of
magnetic structures, especially with the use of neutron scattering, would be
of interest to verify the FOMP scenarios proposed.

When considering the situation in other compounds, we meet with some
problems. The description of FOMP for the systems Pr$_2$Fe$_{14}$B, Nd$_2$Fe$%
_{14}$B should also include non-collinear structures \cite{yamada}, but this
problem is more difficult. In the compound Nd$_2$Fe$_{14}$B a conical
structure occurs in zero field, although the signs of MA for Nd and Fe
sublattices coincide. Usually such a behavior is attributed to the influence
of higher-order anisotropy constants; possibility of different role of
various Nd positions is also discussed. It should be noted that the FOMP
picture is similar for the compound Nd$_2$Co$_{14}$B \cite{ndco} where the
Co ions have easy-plane MA. For the system Sm$_{1-x}$Nd$_x$Co$_5$ \cite{erm}
the RE sublattice already includes two subsystems with opposite MA signs.
However, to obtain FOMP in spirit of our mechanism one still has to divide
this sublattice into two symmetric parts.

The author is grateful to Yu.P. Irkhin, N.V. Mushnikov, A.N. Pirogov, A.S.
Ermolenko and E.V. Rosenfeld for helpful discussions. The work is supported
in part by the Grant of RFFI No. 00-15-96544 (Support of Scientific Schools).

\newpage\ {\sc Figure captions}

Fig.1. Schematic picture of FOMP in a simple toy model with ferromagnetic
exchange $J_{{\rm RT}},$ $K_{{\rm R}}>0$ and large easy-plane $K_{{\rm T}}<0$
(a) the spin configuration in the zero magnetic field (b) the spin
configuration after the transition induced by the magnetic field in the $x$
direction.

Fig.2. Schematic picture of FOMP in the case of antiferromagnetic exchange $%
J_{{\rm RT}},K_{{\rm R}}>0$ (a) the spin configuration in the zero field (b)
the non-collinear spin configuration after the magnetization jump induced by
the magnetic field ${\bf H}\parallel x$.

Fig.3. The field dependence of the total energy in units of $J_{{\rm RT}}=1.$
Other parameter values are $K_{{\rm R}}=0.1,K_{{\rm T}}=-0.15,J_{{\rm RR}%
}=0.5$. Long-dashed line corresponds to the state (a) in Fig.2, and the
short-dashed line to the state (b).

Fig.4. The field dependence of the magnetization in the $x$ direction for
the same parameters as in Fig.3. The observable dependence demonstrating
FOMP (vertical jump) is shown by the solid line.

Fig.5. The magnetization curve in the $x$ direction for $J_{{\rm RT}}=1,K_{%
{\rm R}}=0.1,K_{{\rm T}}=0.1,J_{{\rm RR}}=0.6$.

Fig.6. Schematic picture of FOMP in the case of antiferromagnetic exchange $%
J_{{\rm RT}},$ $K_{{\rm R}}<0.$ The spin configuration is non-collinear
before the jump (a) and becomes collinear after the jump (b).

Fig.7. The magnetization curve in the $x$ direction for $J_{{\rm RT}}=1,K_{%
{\rm R}}=-0.2,K_{{\rm T}}=0.4,J_{{\rm RR}}=0.45$. Long-dashed line
corresponds to the state (a) in Fig.6, and the short-dashed line to the
state (b).

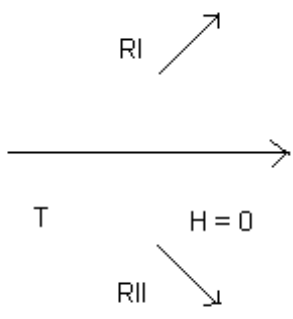
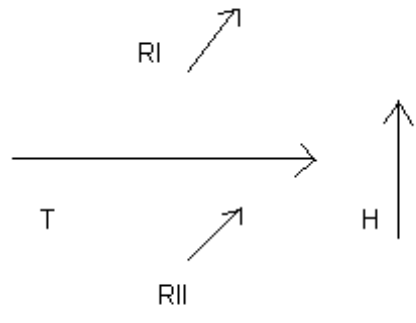
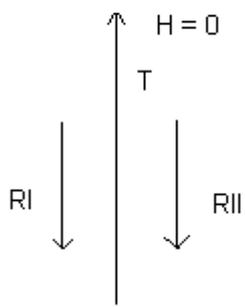
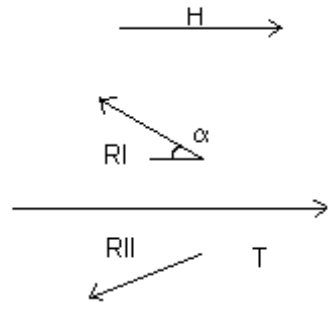
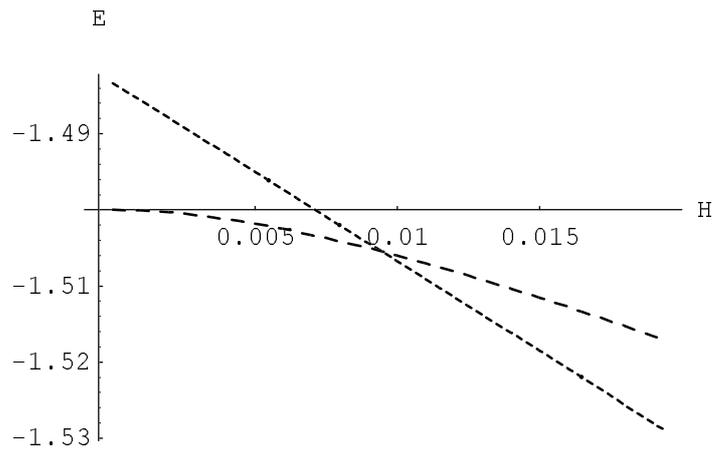
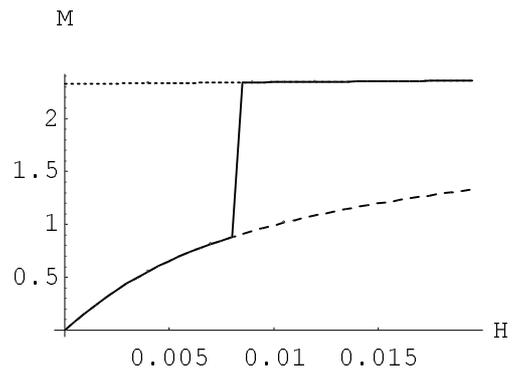
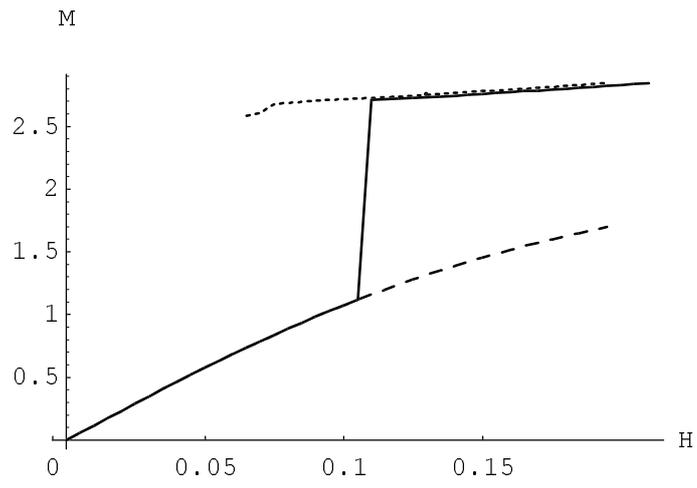
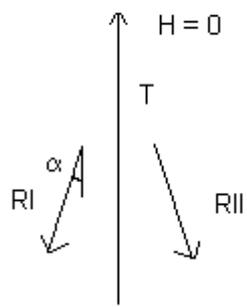
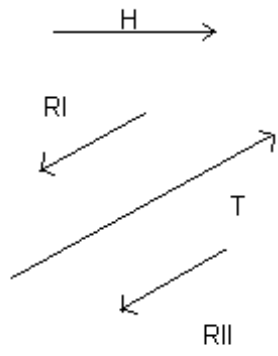
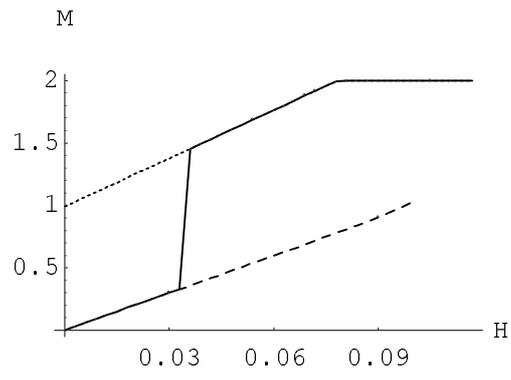
\end{document}